\documentclass[onecolumn,tightenlines,showpacs,nofootinbib,preprintnumbers,
amsmath,amssymb]{revtex4}

\usepackage[usenames,dvipsnames]{color}
\usepackage{caption2}
\usepackage{dcolumn}
\usepackage{graphicx}
\usepackage{bm}
\usepackage{lscape}
\renewcommand{\arraystretch}{0.9}

\usepackage{setspace}
\begin{document}
\begin{spacing}{1.5}


\title{Direct CP violation in $\tau^\pm\rightarrow K^\pm \rho^0 (\omega)\nu_\tau \rightarrow K^\pm \pi^+\pi^-\nu_\tau
$}

\author{Chao Wang$^{1}$\footnote{Email: 201331220003@mail.bnu.edu.cn}, Xin-Heng Guo$^{1}$\footnote{Corresponding author, Email: xhguo@bnu.edu.cn}, Ying Liu$^{1}$\footnote{Email: yingliubnu@gmail.com}, Rui-Cheng Li$^{1}$\footnote{Email: rui-chengli@163.com}
}
\affiliation{\small $^{1}$College of Nuclear Science and Technology, Beijing Normal University, Beijing 100875, China\\
}

\begin{abstract}

We study the direct CP violation in the $\tau^\pm\rightarrow K^\pm \rho^0 (\omega)\nu_\tau \rightarrow K^\pm \pi^+\pi^-\nu_\tau$ decay process in the Standard Model. An interesting mechanism involving the charge symmetry violating mixing between $\rho^0$ and $\omega$ is applied to enlarge the CP asymmetry. With this mechanism, the maximum differential and localized integrated CP asymmetries can reach $-(5.6^{+2.9}_{-1.7})\times10^{-12}$ and $6.3^{+2.4}_{-3.3}\times 10^{-11}$, respectively, which still leave plenty room for CP-violating New Physics to be discovered through this process.

\end{abstract}


\maketitle

\section{Introduction}

CP violation was first observed in the neutral kaon system fifty years ago \cite{Christenson}. The asymmetry of CP in the $K$ meson system can be explained by a weak complex phase in the Cabibbo-Kobayashi-Maskawa (CKM) matrix in the Standard Model (SM) \cite{Cabibbo,Kobayashi}. However, the fundamental origin of CP violation is still an open problem and it is not clear if the CKM mechanism is the only source for CP violation. New Physics (NP) may exist \cite{Huang,Pich,Daiji} and cause CP violation. To verify the origin of CP violation and look for NP, one needs to collect more information about CP violation in as many processes as possible. One such possible process is the $\tau$ decay. $\tau$ is the only lepton which is heavy enough to decay into hadrons and the pure leptonic and semileptonic character of $\tau$ decays provides a clean laboratory to test the structure of the weak currents and the universality of their couplings to the gauge bosons \cite{Pich}. More importantly, with the establishment of the high-luminosity Super $\tau$-Charm factories, more $\tau$ leptons will be produced and its properties will be measured to a very high precision \cite{Huang}. After the CLEO-c experiment ceased data collection in March 2008, the BESIII experiment began to collect data, and the luminosity reached $10^{32}\,\mathrm{cm^{-2}s^{-1}}$ in 2013 \cite{Qin}. Future high luminosity Super $\tau$-Charm factories are also being considered in Russia and Italy and may reach the luminosity of $10^{35}\,\mathrm{cm^{-2}s^{-1}}$ \cite{Peter, E.L.,A.E.,Mikhail}. Moreover, Super $B$-Factories (with the luminosity of $10^{36}\,\mathrm{cm^{-2}s^{-1}}$) will produce about $10^{10}$ $\tau$ pairs per year at the $\Upsilon$(4S) peak \cite{Li,M.B.}. The large statistics collected have considerably improved the statistical accuracy of the $\tau$ measurements and brought a new level of systematic understanding, allowing us to make sensible tests of the $\tau$ properties, provide more information about CP asymmetries in $\tau$ decay processes and seek for the fundamental origin of CP violation.

Experimental searches for CP violating asymmetries in $\tau$ lepton semileptonic decays have been carried out. The missing evidence for a nonzero CP asymmetry was interpreted in terms of a coupling $\Lambda$ in the decay $\tau^\pm \to \pi^\pm \pi^0 \nu_\tau$ \cite{Avery}. Recently, the $\tau^\pm \to K_S \pi^\pm \nu_\tau$ rate asymmetry was measured to be of order $\mathcal{O} (10^{-3})$ by Belle \cite{Bischofberger} and BaBar \cite{Lees}. In order to improve our understanding of CP violation in $\tau$ decays, more efforts should be put on the theoretical side and it is important to study the possibility of finding CP signals in $\tau$ decays. In the framework of the SM, the direct CP asymmetries come about due to a relative weak (CP-odd) and a relative strong (CP-even) phase. This mechanism is forbidden in $\tau$ decays in the leading order of the Fermi coupling constant $G_F$ \cite{Alakabha}. Explicit studies of the decay modes $\tau^\pm\rightarrow K^\pm\pi^+\pi^-\nu_\tau$ \cite{Kilian,Ken}, $\tau^\pm\rightarrow \pi^\pm K^+K^-\nu_\tau$ \cite{Kilian}, $\tau^\pm\rightarrow (3\pi)^\pm\nu_\tau$ \cite{Datta,Choi} and $\tau^\pm\rightarrow (4\pi)^\pm\nu_\tau$ \cite{Datta} show that sizeable CP-violating effects could be generated in some models of CP violation involving several Higgs doublets or left-right symmetry. In order to be sure that any eventual observation of CP violation in $\tau$ decays has its origin beyond the SM, it is essential to study the magnitude of CP violation within the SM.

Usually, vanishingly small CP violation in $\tau$ decays is predicted in the SM. For example, the CP violation in the $\tau^\pm \rightarrow K^\pm \pi^0 \nu_\tau$ mode is estimated to be of order $\mathcal{O} (10^{-12})$ when one takes higher order electroweak corrections into account \cite{Delepine1}. Note that for the decay $\tau^\pm \rightarrow K_S \pi^\pm \nu_\tau$, the SM predicts a CP violating asymmetry of $3.3\times 10^{-3}$ due to the $K_0-\bar{K}_0$ mixing amplitude \cite{Bigi}. In order to obtain a larger CP asymmetry in the SM, one needs to appeal to some phenomenological mechanisms. The charge symmetry violating mixing between $\rho^0$ and $\omega$ ($\rho$-$\omega$ mixing) has been applied in hadron decays for this purpose in the past few years. $\rho $-$ \omega$ mixing has the dual advantages that the strong phase difference is large and well known \cite{Enomoto,Gardner1}. From a series of studies on CP violation, it has already been found that this mechanism can provide a very large strong phase difference (usually 90 degrees) when the mass of the decay product of $\rho^0$($\omega$) , $\pi^+\pi^-$, is in the vicinity of the $\omega$ resonance in some decay channels of heavy hadrons including $B$, $\Lambda_b$, and $D$ \cite{Enomoto,Gardner1,Guo1,Guo2,Leitner}. We will apply this mechanism to the $\tau$ lepton decay in the present paper.

We will consider the decay process $\tau^\pm \to K^\pm \pi^+ \pi^- \nu_\tau$. The CP violation of this process was analyzed theoretically with NP effects in the past \cite{Kilian,Ken}. Now, we investigate the CP violation in this decay mode in the framework of the SM. The interference between the leading order diagram in $G_F$ [Fig.~1(a)] and the second order weak diagrams [Fig.~1(b) and (c)] generates a small CP violation phase \cite{Delepine1}. $\rho $-$ \omega$ mixing has been applied for getting a large strong phase when the invariant mass of the $\pi^+\pi^-$ pair is near the $\omega$ resonance. Hence one can expect that there could be a bigger CP violating asymmetry in the $\tau^\pm\rightarrow K^\pm \rho^0 (\omega)\nu_\tau \rightarrow K^\pm \pi^+\pi^-\nu_\tau$ process. Actually, it will be shown from our explicit calculations that $\rho$-$\omega$ mixing does enlarge the differential CP violating asymmetry by a maximum of four orders of magnitude and the localized integrated CP asymmetry by a maximum of three orders of magnitude. Even though, there is still a large window for studying effects of nonstandard sources of CP violation in experiments.
\begin{figure}
\renewcommand{\captionlabeldelim}{.~}
\centering
\captionstyle{flushleft}
\scalebox{1.0}[1.0]{\includegraphics{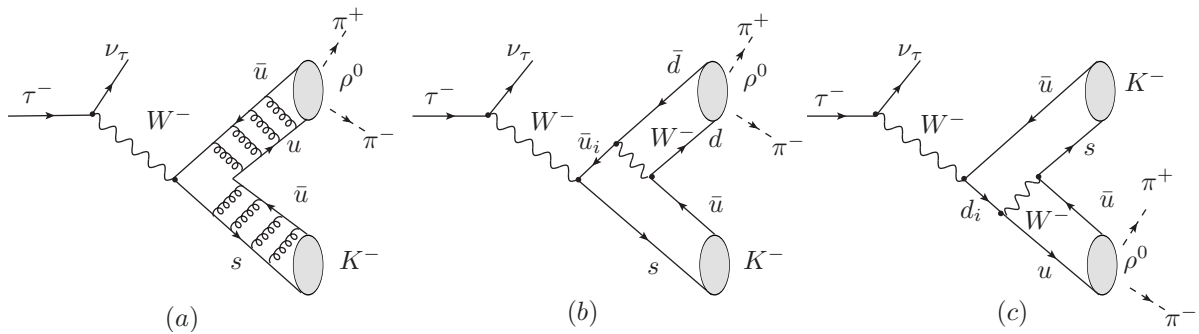}}
\caption{The leading order [(a)] and higher order diagrams [(b) and (c)] in $G_F$ contributing to the decay $\tau^-\to K^-\rho^0\nu_\tau\to K^-\pi^+\pi^-\nu_\tau$. Gluons in (a) are soft ones representing nonperturbative QCD interaction. $u_i=u,\,c,\,t$ in (b) and $d_i=d,\,s,\,b$ in (c).}
\end{figure}

The remainder of this paper is organized as follows. In Sec.~II, we first present the formalism for the CP asymmetry in $\tau^\pm\rightarrow K^\pm \rho^0 (\omega)\nu_\tau \rightarrow K^\pm \pi^+\pi^-\nu_\tau$ via $\rho$-$\omega$ mixing. Then we give the derivation details of the leading order and the second order weak process matrix elements and apply $\rho$-$\omega$ mixing to generate a large CP asymmetry. In Sec.~III, with the expression of meson wave functions and form factors and several parameters we calculate numerical results of the differential and localized integrated CP asymmetries. Our conclusion is included in Sec.~IV.

\section{CP VIOLATION IN $\tau^\pm\rightarrow K^\pm \rho^0 (\omega)\nu_{\tau}\rightarrow K^\pm \pi^+\pi^-\nu_{\tau}$}

A decay process described by some amplitudes may have CP-even and -odd relative phases. Within the SM, the CP-odd relative phase is always a weak phase difference which is directly determined by the CKM matrix. On the contrary, the CP-even relative phase is usually a strong phase difference due to some complicated phenomenological mechanism. Letting $M$ and $\bar M$ be the amplitudes for $\tau^-\rightarrow K^- \rho^0 (\omega)\nu_{\tau}\rightarrow K^- \pi^+\pi^-\nu_{\tau}$ and its CP conjugate one, respectively, we define the two amplitudes as follows:
\begin{eqnarray}
M&=&g_1r_1\mathrm{e}^{\mathrm{i}\phi_1}+g_2r_2\mathrm{e}^{\mathrm{i}\phi_2},\\
\bar M&=&g_1^*r_1\mathrm{e}^{\mathrm{i}\phi_1}+g_2^*r_2\mathrm{e}^{\mathrm{i}\phi_2},
\end{eqnarray}
where $g_1$ and $g_2$ represent CP-odd complex terms which involve coupling constants and CKM matrix elements, $r_1e^{\mathrm{i}\phi_1}$ and $r_2e^{\mathrm{i}\phi_2}$ terms are even under the CP transformation. Then, one has
\begin{eqnarray}
|M|^2-|\bar M|^2&=&4r_1r_2\,\mathfrak{I}\mathfrak{m}(g_1^*g_2)\,\sin(\phi_1-\phi_2)\nonumber\\
&=&4r_1r_2|g_1||g_2|\,\sin [\,\mathrm{Arg}(g_2/g_1)]\,\sin(\phi_1-\phi_2),
\end{eqnarray}
from which, we can see explicitly that both the CP-odd phase difference $\mathrm{Arg}(g_2/g_1)$ and the CP-even phase difference $\phi_1-\phi_2$ are needed to produce CP violation. It will be shown below that the CP-odd phase difference arises from the second order weak processes and the CP-even phase difference is determined by the decay widths of intermediate resonances and $\rho$-$\omega $ mixing in the $\tau^-\rightarrow K^\pm \rho^0 (\omega)\nu_{\tau}\rightarrow K^\pm \pi^+\pi^-\nu_{\tau}$ decay mode.

\subsection{General formalism for CP asymmetry}

The hadronic $\tau$ decay amplitude can be factorized into a purely leptonic part including $\tau$ lepton and neutrino and a hadronic part, where the hadronic system is created from the vacuum via the charged weak current. Thus, the amplitude of $\tau^{-}$ decaying into the $K^-\pi^+\pi^-\nu_{\tau}$ final state through $K^-\rho^0 \nu_\tau$ with the invariant mass of the $\pi^+\pi^-$ pair near the $\rho^0$ resonance can be written in the following general form:
\begin{eqnarray}
 M^\rho&=&\frac{G_{F}}{\sqrt{2}}g_{\rho\pi\pi}s_{\rho}L^{\mu}H^\rho_{\mu},
\end{eqnarray}
where $g_{\rho\pi\pi}$ is the effective coupling for $\rho \rightarrow \pi \pi$, $H^\rho_{\mu}$ is the hadronic matrix element creating $\rho^0K^-$, $L^{\mu}$ is the lepton transition matrix element which can be written as $\bar{u} _{\nu_\tau}\gamma^\mu(1-\gamma^5)u_\tau$ with $u_{\nu_\tau}$ and $u_\tau$ being the Dirac spinors of $\nu_\tau$ and $\tau$, respectively, and $s_\rho$ is the propagator of the $\rho^0$ meson,
\begin{equation}
s_\rho=\frac{1}{s-m^2_\rho+\mathrm{i}m_\rho \Gamma_\rho},
\end{equation}
where $\sqrt{s}$ is the invariant mass of the $\pi^+ \pi^-$ pair, and $m_\rho$ and $\Gamma_\rho$ are the mass and width of the $\rho^0$ meson, respectively. It should be noted that we assume that the $\rho^0$ meson is on-shell since the invariant mass of the $\pi^+\pi^-$ pair is near the mass of the $\rho^0$ meson.

Because of the absence of the CP-odd phase, the CP asymmetry is zero in the leading order in $G_F$ in the SM in the $\tau$ decay. In order to have a nonzero CP violating asymmetry, the second order weak terms corresponding to Fig.~1(b) and (c) (with $u_i=u,\,c,\,t$ and $d_i=d,\,s,\,b$), which provide a CP-odd phase difference, should be taken into account \cite{Delepine1}. The leading order amplitude is denoted by $M_0^\rho$ corresponding to Fig.~1(a) and the second order weak terms are denoted by $M^\rho_{1}$ and $M^\rho_{2}$ corresponding to Fig.~1(b) and (c), respectively.

As mentioned before, in order to obtain a large CP violation, we intend to apply the $\rho $-$ \omega $ mixing mechanism, which leads to large strong phase differences in heavy hadron decays. In this scenario, to the first order of isospin violation, we have the following total amplitude when the invariant mass of the $\pi^+\pi^-$ pair is near the $\omega$ resonance mass:
\begin{eqnarray}
 M=M^\rho+M^{\rho-\omega},
\end{eqnarray}
 with
\begin{eqnarray}
 M^{\rho-\omega}=\frac{G_{F}}{\sqrt{2}}g_{\rho\pi\pi}s_{\rho}L^{\mu}H^{\omega}_{\mu}s_\omega\tilde{\Pi}_{\rho \omega},
\end{eqnarray}
where $\tilde{\Pi}_{\rho \omega}$ is the effective $\rho$-$\omega$ mixing amplitude, $s_\omega$ is the propagator of the $\omega$ meson, and $H_{\mu}^{\omega}$ includes three $\omega K$ annihilation terms $H_{\mu}^{0\omega}$, $H_{\mu}^{1\omega}$ and $H_{\mu}^{2\omega}$ corresponding to Fig.~1(a), (b) and (c), but through the $\omega$ intermediate resonance, respectively. We also assume that the $\omega$ meson is on-shell. It should be noted that the $\rho \to \omega \rightarrow \pi^+ \pi^- $ process has been neglected since it is of the second order of isospin violation. The direct coupling $\omega \rightarrow \pi^+ \pi^- $ has been effectively absorbed into $\tilde{\Pi}_{\rho \omega}$ \cite{O'Connell}. This leads to the explicit $s$ dependence of $\tilde{\Pi}_{\rho \omega}$. Making the expansion $\tilde{\Pi}_{\rho \omega}(s)=\tilde{\Pi}_{\rho \omega}(m^2_\omega)+(s-m_\omega)\tilde{\Pi}_{\rho \omega}^\prime(m^2_\omega)$, the $\rho$-$\omega $ mixing parameters were fitted by Gardner and O'Connell \cite{Gardner}:
\begin{eqnarray}
\mathfrak{R} \mathfrak{e}\tilde{\Pi}_{\rho \omega}(m^2_{\omega})&=&{}-3500\pm 300\ MeV^2,\nonumber\\
\mathfrak{I} \mathfrak{m}\tilde{\Pi}_{\rho \omega}(m^2_{\omega})&=&{}-300\pm 300\ MeV^2,\\
\tilde{\Pi}_{\rho \omega}^\prime(m^2_{\omega})&=&0.03\pm\ 0.04.\nonumber
\end{eqnarray}

We define $M_0=M^\rho_0+M^{\rho-\omega}_0$, $M_1=M_1^\rho+M_1^{\rho-\omega}$ and $M_2=M_2^\rho+M_2^{\rho-\omega}$, where $M_0^{\rho-\omega}$, $M_1^{\rho-\omega}$ and $M_2^{\rho-\omega}$ correspond to $H_{\mu}^{0\omega}$, $H_{\mu}^{1\omega}$ and $H_{\mu}^{2\omega}$, respectively. The CP violation can arise from the interference between $M_0$ and $M_{1}$, $M_{2}$. It should be noted that $M_{1}$ and $M_{2}$ are the second order in $G_{F}$. Therefore, to the $G_F^3$ order, the square of the total amplitude $M=M_0+M_1+M_2$ can be written as
\begin{equation}
 |M|^2=(M_0+M_1+M_2)^{\dagger}(M_0+M_1+M_2)=M_0^{\dagger}M_0+(M_0^{\dagger} M_1 +M_0 M_1^{\dagger})+(M_0^{\dagger} M_2 +M_0 M_2^{\dagger}).
\end{equation}
Then, the differential CP asymmetry which is defined as
\begin{equation}
A_{CP}=\frac {|M|^2-\bar{|M|}^2} {|M|^2+\bar{|M|}^2},
\end{equation}
can be written as the following to the order $G_F$:
\begin{equation}
A_{CP}=\frac{|M|^2-\bar{|M|}^2}{|M_0|^2+\bar{|M_0|}^2},
\end{equation}
where the $M_0^{\dagger} M_1 +M_0 M_1^{\dagger}$ and $M_0^{\dagger} M_2 +M_0 M_2^{\dagger}$ terms are negligible in the denominator since they do not contribute to the second order in $G_{F}$. When we take $\rho$-$\omega$ mixing into account, the three terms in Eq.~(9) can be rewritten in the following forms:
\begin{eqnarray}
M_0M_0^\dagger&=&(G_F/\sqrt{2})^2g_{\rho \pi \pi}^2s_\rho^2 L^{\mu \nu}\big(H_{\mu \nu}^{0\rho 0\rho}+H_{\mu \nu}^{0\rho 0\omega}\tilde{\Pi}^*_{\rho\omega}s_\omega^*+H_{\mu \nu}^{0\omega 0\rho}\tilde{\Pi}_{\rho\omega}s_\omega+H_{\mu \nu}^{0\omega 0\omega}\tilde{\Pi}^2_{\rho\omega}s_\omega^2\big),\nonumber \\
M_0M_1^\dagger &=&(G_F/\sqrt{2})^2g_{\rho \pi \pi}^2s_\rho^2 L^{\mu \nu}\big(H_{\mu \nu}^{0\rho 1\rho}+H_{\mu \nu}^{0\rho 1\omega}\tilde{\Pi}^*_{\rho\omega}s_\omega^*+H_{\mu \nu}^{0\omega 1\rho}\tilde{\Pi}_{\rho\omega}s_\omega+H_{\mu \nu}^{0\omega 1\omega}\tilde{\Pi}^2_{\rho\omega}s_\omega^2\big),\nonumber \\
M_0M_2^\dagger &=&(G_F/\sqrt{2})^2g_{\rho \pi \pi}^2s_\rho^2 L^{\mu \nu}\big(H_{\mu \nu}^{0\rho 2\rho}+H_{\mu \nu}^{0\rho 2\omega}\tilde{\Pi}^*_{\rho\omega}s_\omega^*+H_{\mu \nu}^{0\omega 2\rho}\tilde{\Pi}_{\rho\omega}s_\omega+H_{\mu \nu}^{0\omega 2\omega}\tilde{\Pi}^2_{\rho\omega}s_\omega^2\big),
\end{eqnarray}
where  $L^{\mu\nu}=L^\mu(L^\nu)^\dagger$ and for example, $H^{0\rho0\rho}_{\mu\nu}=H^{0\rho}_\mu(H^{0\rho}_\nu)^\dagger$ .

\subsection{Derivation details of matrix elements}

\begin{figure}
\renewcommand{\captionlabeldelim}{.~}
\centering
\captionstyle{flushleft}
\scalebox{1.0}[1.0]{\includegraphics{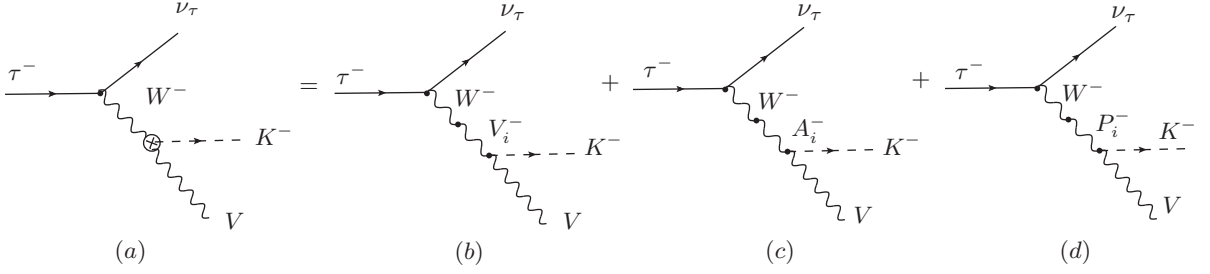}}
\caption{The Feynman diagrams with the intermediate virtual mesons that connect the weak current and the strong vertex in the decay $\tau^- \to VK^-\nu_\tau $ [$V$ is $\rho^0$ (or $\omega$)]. (a) represents the total effective strong vertex; (b), (c) and (d) correspond to the $V_i$ (vector), $A_i$ (axial-vector) and $P_i$ (pseudoscalar) intermediate meson processes, respectively.}
\end{figure}
The transition from the vacuum to the pseudoscalar meson $K^-$ and the vector one $\rho^0$ $(\omega)$ occurs via weak vector and axial-vector current. Based on Lorentz invariance and parity and time-reversal invariance, one can decompose the hadronic matrix element in terms of four form factors in the leading order in $G_F$ \cite{flores}:
\begin{eqnarray}
H_{\mu}^{0\rho(\omega)}&=&-\mathrm{i}V_{us}^*<\rho^{0}(\omega)K^{-}|\overline{s}\gamma_{\mu}(1-\gamma_{5})u|0>\nonumber\\
&=&V_{us}^*\big[-g\varepsilon_{\mu \nu \alpha\beta}\epsilon^{*\nu} p_1^{\alpha}p_2^{\beta}-\mathrm{i}f\epsilon^*_\mu-\mathrm{i}(a_1p_{1\mu}+a_2p_{2\mu})(\epsilon^*\cdot Q)\big],
\end{eqnarray}
where $V_{us}$ is the CKM matrix element, $\bar s$ and $u$ are quark field operators, $p_{1}$ and $p_{2}$ are momenta of $\rho^{0}$ (or $\omega$) and $K^{-}$, respectively, $Q=p_1+p_2$ is the momentum transfer to the hadronic system, $g$ is the vector current form factor, $f$, $a_1$ and $a_2$ are axial-vector current form factors, and $\epsilon _\mu$ denotes the polarization vector of $\rho^0$ (or $\omega$) which satisfies $p_1\cdot\epsilon=0$ and \begin{math}\sum_{\lambda=0,\pm}\epsilon^{*\mu}(q,\lambda)\epsilon ^{\nu}(q,\lambda)=-g^{\mu \nu}+q^{\mu}q^{\nu}/{m_V}\end{math}, where $\lambda=\pm,\,0$ represent the traverse and longitudinal polarizations, respectively, and $m_V$ is the mass of the vector meson $V$ ($V=\rho^0$ or $\omega$). The form factors are functions of $Q^2$ only. They are difficult to be related directly to experimental measurements but can be dealt with in phenomenological models. We will calculate the form factors with the meson dominance model \cite{flores}. The pseudoscalar and vector meson annihilation process in the leading order in $G_F$ is generated by the strong interaction. In the meson dominance model it is assumed that intermediate mesons connect the weak current and the strong vertex shown in the Feynman diagrams in Fig.~2. Using the Feynman rules for these diagrams, the following expressions for the form factors are obtained \cite{flores}:
\begin{eqnarray}
f&=&-\tfrac{1}{2}(Q^2+m_V^2-m_K^2)\sum_i\frac{h_{A_i}t_{A_iVK}}{D_{A_i}(Q^2)}, \qquad g=\tfrac{1}{2}\sum_i\frac{h_{V_i}t_{V_iVK}}{D_{V_i}(Q^2)},\nonumber\\
a_1&=&\tfrac{5}{2}\sum_i\frac{h_{P_i}t_{P_iVK}}{D_{P_i}(Q^2)}+\tfrac{1}{2}\sum_i\frac{h_{A_i}t_{A_iVK}}{D_{A_i}(Q^2)}, \qquad a_2=\tfrac{3}{2}\sum_i\frac{h_{P_i}t_{P_iVK}}{D_{P_i}(Q^2)}+\tfrac{1}{2}\sum_i\frac{h_{A_i}t_{A_iVK}}{D_{A_i}(Q^2)},
\end{eqnarray}
where $V_i$, $A_i$ and $P_i$ denote vector, axial-vector and pseudoscalar intermediate meson resonances, respectively, $h_{M_i}$ ($M_i=A_i$, $V_i$ or $P_i$) denotes the weak coupling of the $M_i$ intermediate meson, $t_{M_iVK}$ is its strong coupling to the $VK$ final state, $m_K$ is the mass of the $K$ meson, and $D_{M_i}\equiv Q^2-m_{M_i}^2+im_{M_i} \Gamma_ {M_i}$ where $m_{M_i}$ ($\Gamma_{M_i}$) is the mass (width) of the corresponding intermediate meson. The details about the intermediate mesons and their weak couplings and strong vertex coupling constants will be given in Section III. From Eqs.~(12) and (13) it can be found that in the leading order in $G_F$ the CP-odd phase is absent, and the CP-even phase is determined by the decay widths of intermediate resonances when $\rho$-$\omega$ mixing is not considered.

Next, we proceed to evaluate $M_1$ and $M_2$ based on the perturbation method. Note that $M_0M_2^\dagger+M_0^\dagger M_2$ in Eq.~(9) is proportional to $|V_{ud_i}|^2|V_{us}|^2$ and will not contribute to CP violation. Hence we only have to consider $M_1$. In the framework of perturbation method, it can be evaluated in a similar way to $B$ decays \cite{Du}. Since the $\tau$ mass is much smaller than the $W$-boson mass $M_W$, the momenta of all the particles involved in the $\tau$ decay are much smaller than $M_W$. As a result, we can approximate the denominator of the $W$-boson propagator $(p_1+p_2)^2-M_W^2$ by $-M^2_W$ in the numerator of the $W$-boson propagator. The wave functions including spin factors of pseudoscalar and vector mesons are taken as \cite{Falk}
\begin{eqnarray}
\Psi_V(x,p)&=&{}-\frac{I}{2\sqrt{2}\sqrt{3} m_V} \phi_V(x)(m_V+p\!\!\!/)\epsilon \!\!\!/,\\
\Psi_P(x,p)&=&{}-\frac{I}{2\sqrt{2}\sqrt{3} m_P} \phi_P(x)(m_P+p\!\!\!/)\gamma_5,
\end{eqnarray}
where $I=3$ is an identity in color space, $m_P$ and $m_V$ are the masses of the pseudoscalar and vector mesons, respectively, $p$ represents the momentum of the meson $P$ or $V$, $x$ is the longitudinal momentum fraction of the constituent quark, and the non-perturbution effects are included in the distribution amplitudes $\phi_V(x)$ and $\phi_P(x)$, which satisfy $\int^1_0 \phi_{V(P)}(x)\mathrm{d}x=f_{V(P)}/(2\sqrt{6})$, where $f_{V(P)}$ is the decay constant of $V(P)$. According to the Feynman diagram (b) in Fig.~1, the hadronic matrix elements $H^{1\rho}$ (or $H^{1\omega}$) can be expressed as
\begin{eqnarray}
H^{1\rho(\omega)}_{\mu}&=&+(-)\frac{G_F}{\sqrt{2}}(2\pi )^3I\sum_{i}V^*_{u_i s}V_{u_i d}V^*_{ud}\sqrt{m_u m_d^2 m_s}\int_0^1\mathrm{d}x\mathrm{d}y\phi^*_{\rho(\omega)}(x)\phi^*_{K}(y) \nonumber\\
&&\qquad \cdot \frac{1}{2\sqrt{2} m_{\rho(\omega)}}(m_{\rho(\omega)}-p\!\!\!/_2)\epsilon \!\!\!/^*\gamma _{\alpha}(1-\gamma _5)\frac {1}{2\sqrt{2}m_K}(m_K-p\!\!\!/_1)\gamma _5 \gamma _\mu(
1-\gamma _5)I_{u_i}\gamma ^ \alpha (1-\gamma _5),
\end{eqnarray}
where $V_{u_i s}$ and $V_{u_i d}$ are the CKM matrix elements, $+(-)$ corresponds to $\rho(\omega)$ and we define $I_{u_i}=\text{i}(p\!\!\!/_{u_i}+m_{u_i})/(p_{u_i}^2$\\$-m_{u_i}^2)$ with $m_{u_i}$ and $p_{u_i}$ being the current quark mass and the momentum of the intermediate quark $u_i$, respectively. We will neglect the difference between the masses of $\rho^0$ and $\omega$ mesons in the following, i.e., we take $m_\rho=m_\omega$.

Using the unitarity of the CKM matrix, we have
\begin{eqnarray}
\sum_{i}V^*_{u_i s}V_{u_id}I_{u_i}&=&V^*_{ud}V_{us}(I_u-I_c)+V^*_{td}V_{ts}(I_t-I_c)\nonumber\\
&\approx &V^*_{ud}V_{us}(I_u-I_c)-V^*_{td}V_{ts}I_c
\end{eqnarray}
where the last line is obtained using the fact that $m_t$ is much larger than masses of other quarks involved in this process. We note that only $V_{td}^*V_{ts}$ provides a weak CP-violation phase, so it is unnecessary to consider the contribution of the first term. As a consequence, the CP asymmetry only depends on $V^*_{td}V_{ts}I_c$. We define
\begin{eqnarray}
A_1&\equiv& \int^1_0\mathrm{d}x\mathrm{d}y\phi^*_{\rho(\omega)}(y)\phi^*_K(x)\frac{1}{xQ^2+(1-x)m_\rho^2+(x^2-x)m_K^2-m^2_{c}},\nonumber\\
&=&\int^1_0\mathrm{d}x\phi^*_K(x)\frac{f_\rho}{2\sqrt{6}[xQ^2+(1-x)m_\rho^2+(x^2-x)m_K^2-m^2_{c}]},\\
B_1&\equiv& \int^1_0\mathrm{d}x\mathrm{d}y\phi^*_{\rho(\omega)}(y)\phi^*_K(x)\frac{x}{xQ^2+(1-x)m_\rho^2+(x^2-x)m_K^2-m^2_{c}}\nonumber\\
&=&\int^1_0\mathrm{d}x\phi^*_K(x)\frac{xf_\rho}{2\sqrt{6}[xQ^2+(1-x)m_\rho^2+(x^2-x)m_K^2-m^2_{c}]}.
\end{eqnarray}
Inserting Eqs.~(18), (19) and (20) into Eq.~(17) and only considering the CP asymmetry term, $H^{1\rho (\omega)}_{\mu}$ can be simplified as
\begin{eqnarray}
H^{1\rho (\omega)}_{\mu}&=& \frac{6\sqrt{2}(2\pi)^3\sqrt{m_u m_d^2 m_s}G_F V_{ts}^*V_{td}V_{ud}^*}{m_k}\nonumber\\
&&\; {}\cdot \Big\{{}-A_1\varepsilon_{\mu \nu \alpha \beta }\epsilon^{*\nu}p_1^{\alpha}p_{2}^{\beta}-\mathrm{i}\epsilon ^{*\mu}\Big[\frac{1}{2}A_1(Q^2-m_\rho^2-m_K^2)+B_1m^2_K\Big]\nonumber\\[6pt]
&&\qquad {}+\mathrm{i}A_1p_1^{\mu}(Q\cdot \epsilon ^*)+\mathrm{i}2B_1p_2^{\mu}(Q\cdot \epsilon ^*)\Big\}.
\end{eqnarray}
We can see that the weak phase appears but the strong phase is absent in this amplitude if $\rho$-$\omega$ mixing is not included.

Now we take $\rho$-$\omega$ mixing into account and show how $\rho$-$\omega$ mixing enlarges the CP violation. In the meson dominance model, the form factors of the annihilation process are dominated by strong interaction. So, we adopt the same form factors in the $K\rho$ and $K\omega$ annihilation processes. According to Eq.~(13), we have $H_{\mu}^{0\rho}=H_{\mu}^{0\omega}$. $H_{\mu}^{1\rho(\omega)}$ is dependent on the hadronic wave functions. Since the wave functions of mesons are determined by strong interaction, which preserves isospin, we assume that the $\rho^0$ and $\omega$ mesons have the same hadronic wave functions. Therefore, from Eq.~(17), we have $H^{1\rho}_{\mu}=-H^{1\omega}_{\mu}$. Then, the first two equations of Eq.~(12) can be written as
\begin{eqnarray}
M_0M_0^\dagger &=&(G_F /\sqrt{2})^2 g_{\rho \pi \pi}^2 s_\rho ^2 L^{\mu \nu}H^{0\rho0\rho}_{\mu\nu}\big(1+\tilde{\Pi}^*_{\rho\omega}s^*_\omega+\tilde{\Pi}_{\rho\omega}s_\omega+\tilde{\Pi}^2_{\rho\omega}s_{\omega}^2\big)\nonumber\\
M_0M_1^\dagger &=&(G_F/\sqrt{2})^2g_{\rho \pi \pi}^2s_\rho^2 L^{\mu \nu}H^{0\rho1\rho}_{\mu\nu}\big(1-\tilde{\Pi}^*_{\rho\omega}s^*_\omega+\tilde{\Pi}_{\rho\omega}s_\omega-\tilde{\Pi}^2_{\rho\omega}s_{\omega}^2\big).
\end{eqnarray}
We can see explicitly that $\rho$-$\omega$ mixing provides additional complex terms to $A_{CP}$. As will be shown later, these complex terms enlarge the CP-even phase, which leads to a bigger CP asymmetry.

Finally, we will calculate $L^{\mu\nu}H^{0\rho0\rho}_{\mu\nu}$ and $L^{\mu\nu}H^{0\rho1\rho}_{\mu\nu}$. For simplicity, we will consider the unpolarized $\tau$ decay process. The unpolarized leptonic scattering tensor is
\begin{eqnarray}
L^{\mu\nu}&=&\tfrac{1}{2}\sum_{\lambda_3,\,\lambda_4} tr \big[\bar{u}_{\nu_{\tau}}(p_3,\,\lambda_3)\gamma^\mu(1-\gamma_5)u_{\tau}(p_4,\,\lambda_4)\bar{u}_\tau\gamma ^{\nu}(p_4,\,\lambda_4)(1-\gamma_5)u_{\nu_\tau}(p_3,\,\lambda_3)\big]\nonumber \\
&=&4\big[{}-g^{\mu \nu}(p_3 \cdot p_4)+p_3^{\mu}\cdot p_4^{\nu}+p_3^{\nu}\cdot p_4^{\mu}+\mathrm{i}\varepsilon ^{\mu \nu \alpha \beta}p_{3 \alpha }p_{4 \beta }\big],
\end{eqnarray}
where $p_3$ and $p_4$ represent the momenta of $\nu_\tau$ and $\tau$, respectively, and $\lambda_3$ and $\lambda_4$ represent the helicities of $\nu_\tau$ and $\tau$, respectively. We also sum over the spins of hadrons. Then, from Eqs.~(13) and (23), one has
\begin{eqnarray}
L^{\mu\nu}H^{0\rho0\rho}_{\mu\nu}&=&4|V_{us}|^2 \big[(-2x_0-m_\rho^2 x_1-m^2_Kx_2)(p_3 \cdot p_4)+2x_1(p_1 \cdot p_3)(p_1 \cdot p_4)+2x_2(p_2 \cdot p_3)(p_2 \cdot p_4)\nonumber\\
&&{}\ \quad\qquad-(x_++x_-)(p_1 \cdot p_2)(p_3 \cdot p_4)+(x_++x_-+2gf^*+2g^*f)(p_1 \cdot p_3)(p_2 \cdot p_4)\nonumber\\
&&{}\ \quad\qquad+(x_++x_--2gf^*-2g^*f)(p_1 \cdot p_4)(p_2 \cdot p_3)\big],
\end{eqnarray}
where
\begin{eqnarray}
x_0&=&-\tfrac{1}{4}g^2(Q^4+m_\rho^4+m_K^4-2m_\rho^2Q^2-2m_K^2Q^2-2m_\rho^2m_K^2)-f^2,\nonumber\\
x_1&=&g^2m_K^2+a_1^2\Big[-Q^2+\tfrac{(p_1 \cdot Q)^2}{m_\rho^2}\Big]+\tfrac{f^2}{m_\rho^2}+\tfrac{p_1 \cdot p_2}{m_\rho^2}fa_1^*+\tfrac{p_1 \cdot p_2}{m_\rho^2}f^*a_1,\nonumber\\
x_2&=&g^2m_\rho^2+a_2^2\Big[-Q^2+\tfrac{(p_1 \cdot Q)^2}{m_\rho^2}\Big]-(a_2 f^*+a_2^* f),\nonumber\\
x_+&=&-\tfrac{1}{2}g^2(Q^2-m_K^2-m_\rho^2)+a_1a_2^*\Big[-Q^2+\tfrac{(p_1 \cdot Q)^2}{m_\rho^2}\Big]-a_1f^*+\tfrac{p_1 \cdot p_2}{m_\rho^2}fa_2^*,\nonumber\\
x_-&=&-\tfrac{1}{2}g^2(Q^2-m_K^2-m_\rho^2)+a_1^*a_2\Big[-Q^2+\tfrac{(p_1 \cdot Q)^2}{m_\rho^2}\Big]-a_1^*f+\tfrac{p_1 \cdot p_2}{m_\rho^2}f^*a_2.\nonumber\\
\end{eqnarray}
From Eqs.~(21) and (23), one has
\begin{eqnarray}
L^{\mu\nu}H^{0\rho1\rho}_{\mu\nu}&=&\frac{6\sqrt{2}(2\pi)^3\sqrt{m_u m_d^2 m_s}G_F V_{ts}V^*_{td}V_{ud}V_{us}^*}{m_K}\nonumber\\
 &&\cdot \big[(-2x^\prime_0- x^\prime_1m^2_\rho-x_2^\prime m^2_K)(p_3 \cdot p_4)+2x^\prime_1(p_1 \cdot p_3)(p_1 \cdot p_4)+2x^\prime_2(p_2 \cdot p_3)(p_2 \cdot p_4)\nonumber\\
&&\ {}+(x^\prime_++x^\prime_-+2fB_1+2g\lambda)(p_1 \cdot p_3)(p_2 \cdot p_4)+(x^\prime_++x^\prime_--2f^*B_1-2g^*\lambda)(p_1 \cdot p_4)(p_2 \cdot p_3)\nonumber\\
&&\ {}-(x^\prime_++x^\prime_-)(p_1 \cdot p_2)(p_3 \cdot p_4)\big],
\end{eqnarray}
where
\begin{eqnarray}
x_0^\prime&=&-\tfrac{1}{4}gA_1(Q^4+m_\rho^4+m_K^4-2m_\rho^2Q^2-2m_K^2Q^2-2m_\rho^2m_K^2)-f\lambda,\nonumber\\
x_1^\prime&=&-gA_1m_K^2+\tfrac{1}{2}gB_1(Q^2-m_\rho^2-m_K^2)-a_1A_1\Big[-Q^2+\tfrac{(p_1\cdot Q)^2}{m_\rho^2}\Big]-A_1f\tfrac{p_1\cdot p_2}{m_\rho^2}+a_1\lambda\tfrac{p_1\cdot p_2}{m_\rho^2},\nonumber\\
x_2^\prime&=&-gA_1m_\rho^2-2a_2B_1\Big[-Q^2+\tfrac{(p_1\cdot Q)^2}{m_\rho^2}\Big]+2B_1f+a_2\lambda,\nonumber\\
x_+^\prime&=&\tfrac{1}{2}gA_1(Q^2-m_K^2-m_\rho^2)-gA_1m_K^2-2B_1f\tfrac{p_1\cdot p_2}{m_\rho^2}-a_1\lambda,\nonumber\\
x_-^\prime&=&\tfrac{1}{2}gA_1(Q^2-m_K^2-m_\rho^2)-a_2A_1\Big[-Q^2+\tfrac{(p_1 \cdot Q)^2}{m_\rho^2}\Big]-A_1f+a_2\lambda\tfrac{p_1\cdot p_2}{m_\rho^2},\nonumber\\
\lambda &=&\tfrac{1}{2}A_1(Q^2-m^2_K-m_\rho^2)+B_1m^2_K.\nonumber
\end{eqnarray}

\subsection{Hadronic rest-frame}

\begin{figure}
\renewcommand{\captionlabeldelim}{.~}
\centering
\captionstyle{flushleft}
\scalebox{1.0}[1.0]{\includegraphics{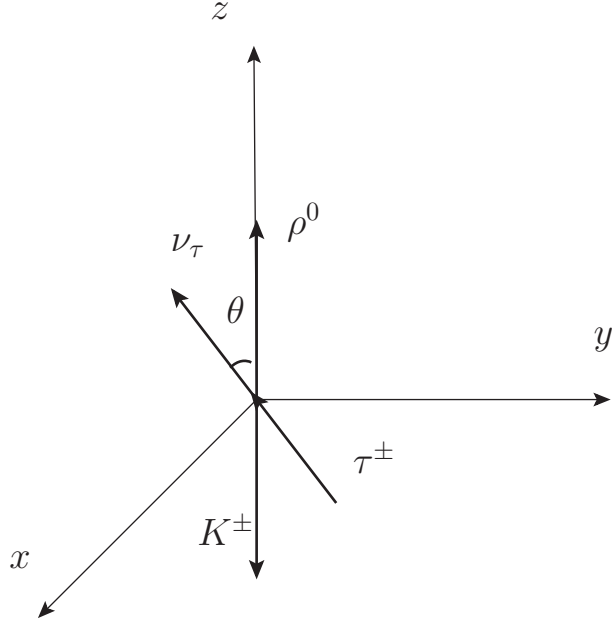}}
\caption{The hadronic rest-frame. The $z$ axis is chosen to be in the direction of the motion of the $\rho^0$ (or $\omega$) meson. The three-momentum of $K$ is chosen to be $\bm{p_2}=-\bm{p_1}$. The $(x, z)$ plane is aligned with the $\rho^0$ and $\nu_\tau$ movement plane, with $\bm{n_\perp=(p_1\times p_3)/|p_1\times p_3|}$ (the normal to the $\rho^0$ and $\nu_\tau$ movement plane) pointing along the $y$ axis. The distribution angle $\theta$ is the one between the motion direction of $\rho^0$ (or $\omega$) and the neutrino.}\label{fig}
\end{figure}

In the previous subsections, we have given the the general expression of CP asymmetry and derivations of matrix elements. For simplicity, we choose to work in a special reference frame and express the products among vectors $p_1$, $p_2$, $p_3$, $p_4$, $Q$ and $\epsilon$ in terms of the square of momentum transfer $Q^2$, the invariant mass of the $\pi^+\pi^-$ pair $\sqrt{s}$, and a distribution angle $\theta$ in this subsection. We note that it is convenient to express the momenta of hadrons and leptons and calculate various components of the matrix elements in the hadronic rest-frame \cite{Kuhn}. This frame is defined in Fig.~\ref{fig}. The $z$ axis is chosen to be in the direction of motion of the $\rho^0$ (or $\omega$) meson. The three-momentum of $K$ is chosen to be $\bm{p_2}=-\bm{p_1}$. The $(x, z)$ plane is aligned with the $\rho^0$ and $\nu_\tau$ movement plane, with $\bm{n_\perp=(p_1\times p_3)/|p_1\times p_3|}$ (the normal to the $\rho^0$ and $\nu_\tau$ movement plane) pointing along the $y$ axis. The distribution angle $\theta$ is the one between the motion direction of $\rho^0$ (or $\omega$) and the neutrino. Then, the momenta of hadrons and leptons in this hadronic rest frame are given as follows:
\begin{eqnarray}
p_1^\mu&=&(E_1,\ 0,\ 0,\ P),\nonumber\\
p_2^\mu&=&(E_2,\ 0,\ 0,\ -P),\nonumber\\
p_3^\mu&=&(K,\ K\sin\theta,\ 0,\ K\cos\theta),\nonumber\\
p_4^\mu&=&(E_4,\ K\sin\theta,\ 0,\ K\cos\theta),\nonumber\\
Q^\mu&=&(E_1+E_2,\ 0,\ 0,\ 0)\nonumber\\
&=&(E_4-K,\ 0,\ 0,\ 0),
\end{eqnarray}
and the polarization vectors of $\rho^0$ (or $\omega$) in this hadronic rest frame are
\begin{eqnarray}
\epsilon^{\lambda=\pm1}&=&(0,\ 1,\ \pm \mathrm{i}\ ,0),\nonumber\\
\epsilon^{\lambda=0}&=&\frac{1}{\sqrt{m_\rho}}(P,\ 0,\ 0,\ E_1),
\end{eqnarray}
with \begin{eqnarray}
E_1&=&\frac{Q^2+m_K^2-m^2_\rho}{2\sqrt{Q^2}},\qquad E_2=\frac{Q^2-m_K^2-m^2_\rho}{2\sqrt{Q^2}},\nonumber\\
P&=&\frac{\sqrt{m_K^4+m^4_\rho-2m_K^2Q^2-2m_K^2m^2_\rho-2Q^2m^2_\rho}}{2\sqrt{Q^2}},\nonumber\\
K&=&\frac{m_\tau^2-Q^2}{2\sqrt{Q^2}},\qquad \qquad E_4=\frac{m_\tau^2+Q^2}{2\sqrt{Q^2}}.
\end{eqnarray}

The above expressions for various hadron and lepton momentum vectors allow us to determine simple expressions for matrix elements which involve products including $p_1 \cdot p_2$, $p_1\cdot p_3$, $p_1\cdot p_4$, $p_2 \cdot p_3$, $p_2 \cdot p_4$, $p_3 \cdot p_4$, $p_1 \cdot Q$, $p_2 \cdot Q$,  $p_3 \cdot Q$, $p_4 \cdot Q$, and $Q\cdot \epsilon$ in the term of $Q^2$, $\sqrt{s}$ and $\theta$. We will integrate over the angle $\theta$ since we will not consider the angle distribution. Furthermore, by integrating $A_{CP}$ in the region $\Omega$ in which $Q^2$ and $s$ vary in some areas, we obtain the localized integrated CP asymmetry which takes the following form:
\begin{equation}
A_{CP}^\Omega=\frac{\int_\Omega \mathrm{d}Q^2\mathrm{d}s(|M|^2-\bar{|M|}^2)}{\int_\Omega \mathrm{d}Q^2\mathrm{d}s(|M_0|^2+\bar{|M_0|}^2)}.
\end{equation}

\section{Numerical results}

From the above discussions, the CP violating asymmetries depend on the values of $Q^2$ and $s$. In this section we give the explicit expressions of meson wave functions and form factors, and values of several parameters in order to calculate the CP violating asymmetries. We find that significant cancellation occurs as one performs the integration over $Q^2$. To show more details about this cancellation, we calculate both the differential and the integrated CP asymmetries. We also compare CP asymmetries with and without $\rho$-$\omega$ mixing.

\subsection{Models for form factors and meson wave functions}

The hadronic $\tau$ decay is dominated by the meson annihilation diagram, Fig.~1(a). As mentioned before, the vector and pseudoscalar meson annihilation form factors in this decay mode are difficult to be related directly to experimental measurements. One therefore needs to adopt phenomenological models. Following Ref.~\cite{flores} we use the meson dominance model in our calculation. In this model it is assumed that the vector form factor $g$ is dominated by the $K^*$(892) and $K^*(1410)$ vector mesons and $f$ and $a_\pm$ are dominated by the exchange of the $K^-$ pseudoscalar meson and the $K_1(1270)$ and $K_1(1400)$ axial-vector mesons \cite{flores}. The expressions for the form factors are given in Eq.~(14). In Ref.~\cite{flores}, the values of weak couplings and strong vertex couplings were extracted from experiments and fixed by the SU(3) flavor symmetry. We display these values in Table~I.

\begin{table}[htb]
\renewcommand{\captionlabeldelim}{.~}
\renewcommand{\arraystretch}{1.3}
\caption{The values of $h_{M_i}$, $t_{M_iVK}$, $m_{M_i}$ and $\Gamma_{M_i}$in the numerical calculations.}
\begin{tabular*}{14.5cm}{@{\extracolsep{\fill}}lr@{$\pm$}lr@{$\pm$}lr@{$\pm$}lr@{$\pm$}lr@{$\pm$}l}
\hline
\hline
&\multicolumn{2}{c}  {Pseudoscalar}  & \multicolumn{4}{c}{Axial Vector} & \multicolumn{4}{c}{Vector} \\
Intermediate mesons & \multicolumn{2}{c} {$K^-$}  &  \multicolumn{2}{c} {$K_1(1270)$}  & \multicolumn{2}{c} {$K_1(1400)$}  & \multicolumn{2}{c}{$K^*(892)$}   & \multicolumn{2}{c}{$K^*(1680)$}  \\
\hline
$h_{M_i}$ ($10^{3}\,\mathrm{MeV}^2$)   &  0.159&0.0015\,$\mathrm{MeV}^{-1}$  & 215&25     & 170&130        &    188&4         & 242&25       \\
$t_{M_iVK}$ ($10^{-3}\,\mathrm{MeV}^{-1}$)  & $-3170$&30\,$\mathrm{MeV}$  & $-1.94$&0.10&  0.48&0.24     & 8.71&0.95        & $-3.71$&2.60 \\
$m_{M_i}$ (MeV)        &   494&0.016                  &  1272&7      &    1403&7      &    892&0.26      &   1717&27    \\
$\Gamma_{M_i}$  (MeV)  &   \multicolumn{2}{c}0        &   90&20      &     174&13     &    50.8&0.9      &    322&110   \\
\hline
\hline
\end{tabular*}
\end{table}

We use the $K$ meson wave function of the Brodsky-Huang-Lepage prescription which have the following form \cite{Guo}:
\begin{eqnarray}
\Phi_K(x,\ \bm{k}_\perp)&=&A_K (1-2x)^2\,\mathrm{exp}\Big[-b_K^2\Big(\frac{\bm k^2_\perp +m_s^ {\prime2}}{x}+\frac{\bm{k}^2_\perp +m_u ^ {\prime2}}{1-x}\Big)\Big],
\end{eqnarray}
where $\bm{k}_\perp$ is the transverse momentum of the constituents of $K$, $m_u^\prime$ and $m_s^\prime$ are the constituent quark masses of $u$ and $s$, respectively. Integrating $\Phi_K(x,\ \bm{k}_\perp)$ over $\bm{k}_\perp$ one has the following distribution amplitude:
\begin{eqnarray}
\phi_K(x)&=&\frac{A_K}{16 \pi^2 b_K^2} x(1-x)(1-2x)^2\,\mathrm{exp}\Big[-b_K^2\Big(\frac{m_s^ {\prime2}}{x}+\frac{m_u ^ {\prime2}}{1-x}\Big)\Big].
\end{eqnarray}
In the following numerical calculations we use the parameters $A_K=232$\,$\mathrm{GeV}^{-1}$, $b_K^2=0.61$\,$\mathrm{GeV}^{-2}$, $m^\prime_u=350$\,MeV, $m_s^\prime=550$\,MeV and $f_\rho=221$\,MeV \cite{Guo}.

\subsection{Numerical results for the CP asymmetries}

We are now ready to evaluate numerical results of CP asymmetries. We take the meson masses $m_\rho=770$\,MeV and $m_K=493$\,MeV, the lepton mass $m_\tau=1776$\,MeV, the current quark masses $m_u=2.3$\,MeV, $m_d=4.8$\,MeV, $m_s=95$\,MeV and  $m_c=1275$\,MeV \cite{Beringer}. The CKM matrix, which elements are determined from experiments, can be expressed in terms of the Wolfenstein parameters $A$, $\rho$, $\lambda$ and $\eta$ \cite{Beringer}:
\begin{equation}
\left(
\begin{array}{ccc}
  1-\tfrac{1}{2}\lambda^2   & \lambda                  &A\lambda^3(\rho-\mathrm{i}\eta) \\
  -\lambda                 & 1-\tfrac{1}{2}\lambda^2   &A\lambda^2 \\
  A\lambda^3(1-\rho-\mathrm{i}\eta) & -A\lambda^2              &1\\
\end{array}
\right),
\end{equation}
where $\mathcal{O} (\lambda^{4})$ corrections are neglected. The latest values for the parameters in the CKM matrix are \cite{Beringer}:
\begin{eqnarray}
\lambda&=&0.22535\pm0.00065,\qquad \ A=0.811^{+0.022}_{-0.012},\nonumber\\
\bar{\rho}&=&0.131^{+0.0026}_{-0.013},\qquad \qquad \quad \bar{\eta}=0.345^{+0.013}_{-0.014},
\end{eqnarray}
with
\begin{eqnarray}
\bar{\rho}=\rho(1-\tfrac{\lambda^2}{2}),\qquad \bar{\eta}=\eta(1-\tfrac{\eta^2}{2}).
\end{eqnarray}

\begin{figure}[htbp]
\renewcommand{\captionlabeldelim}{.~}
\centering
\captionstyle{flushleft}
\scalebox{0.6}{\includegraphics[21,277][610,610]{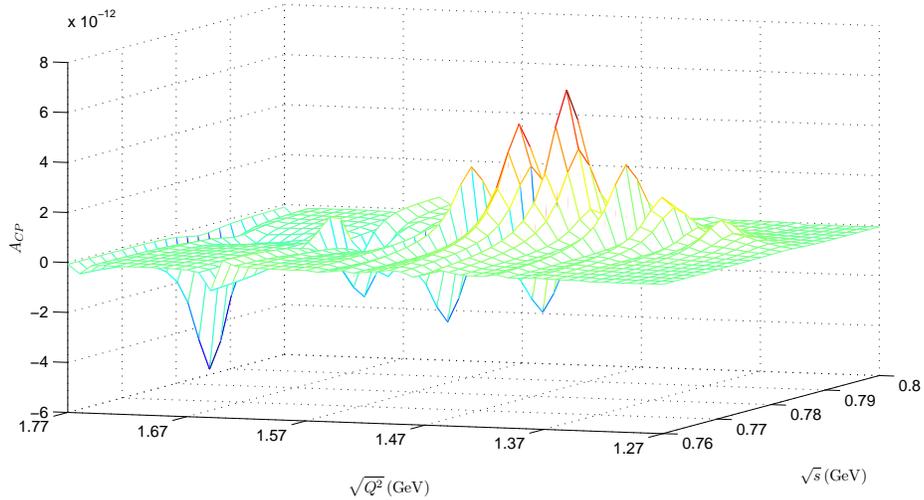}}
\caption{The differential CP asymmetry as a function of $\sqrt s$ and $\sqrt {Q^2}$. The numerical results correspond to central values of the parameters involved in the calculation.}\label{dif}
\end{figure}

In our numerical calculations, the most uncertain factors come from the CKM matrix elements and the form factors in the leading order in $G_F$. In fact, the uncertainties due to the CKM matrix elements are mostly from $\eta$ since $\lambda$ is well determined and the CP violating asymmetries are independent of $\rho$. Hence in the following we take the central value of $\lambda$, 0.225. In the meson dominance model, the uncertainties arising from form factors are dominated by those of the strong and weak coupling constants of the $K_1(1400)$ meson due to the poor quality of measurements. The values of $\rho$-$\omega$ mixing paraments also bring some uncertainties.

\begin{figure}[bt]
\renewcommand{\captionlabeldelim}{.~}
\centering
\captionstyle{flushleft}
\scalebox{0.8}{\includegraphics[8,241][605,695]{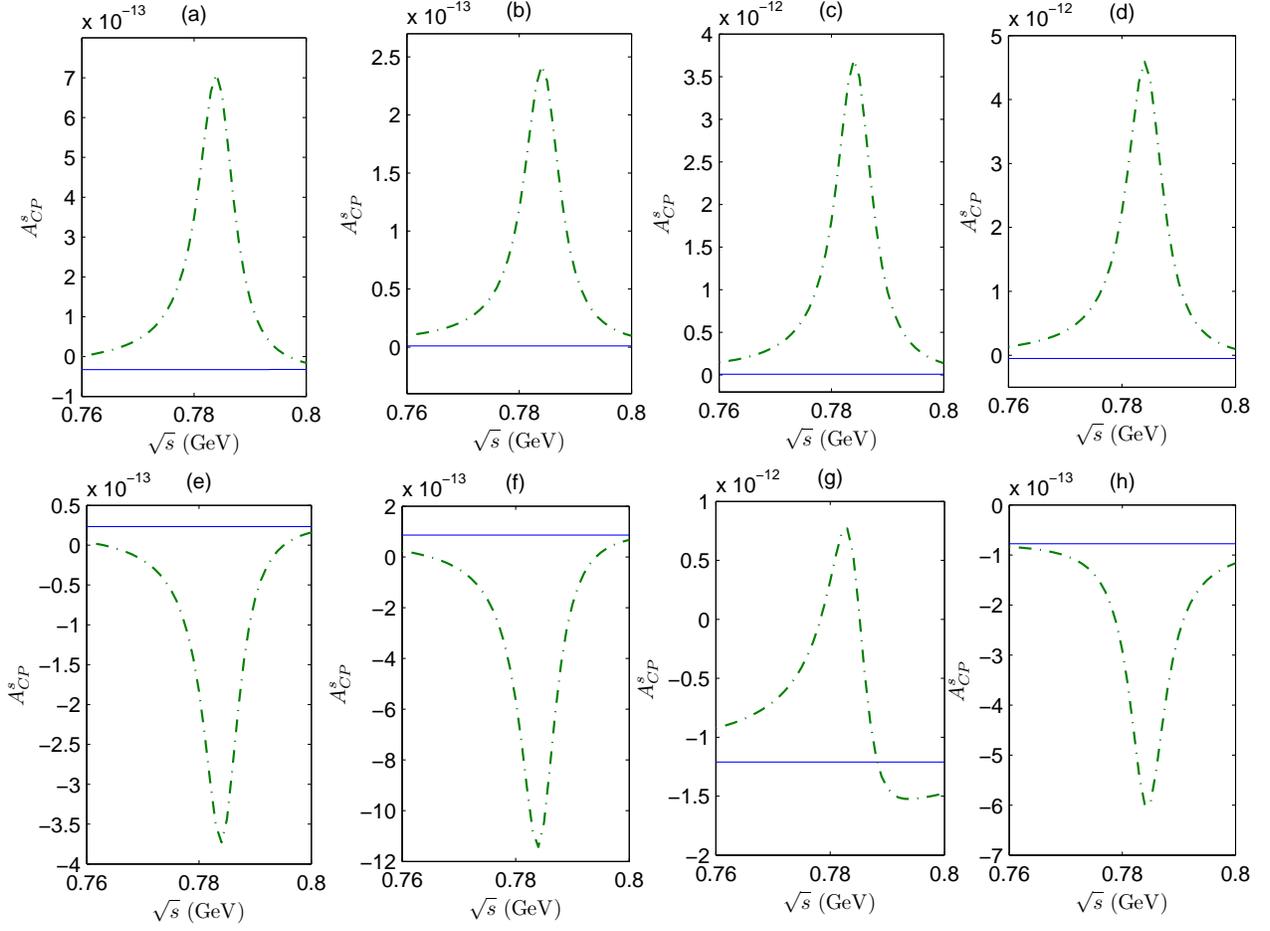}}
\caption{The localized integrated CP asymmetry $A^s_{CP}$ as a function of $\sqrt{s}$. (a) For integrating over $Q^2$ in $\sqrt{Q^2}=$(1.30$\,\text {GeV}$, 1.35$\,\text {GeV}$): the dash-dotted line corresponds to the CP asymmetry including $\rho$-$\omega$ mixing and the solid line corresponds to the CP asymmetry without $\rho$-$\omega$ mixing; (b), (c), (d), (e), (f), (g) and (h) correspond to the integration intervals (1.35 GeV, 1.40 GeV), (1.40 GeV, 1.45 GeV), (1.45 GeV, 1.50 GeV), (1.50 GeV, 1.55 GeV), (1.55 GeV, 1.60 GeV), (1.60 GeV, 1.65 GeV) and (1.65 GeV, 1.70 GeV), respectively. We take central values of the parameters involved in the calculation.}
\label{s}
\end{figure}

In order to find the details about the dependence of the CP violating asymmetries on $Q^2$ and $s$, we study the differential CP asymmetries. Since CP asymmetries are calculated around the $\omega(782)$ resonance region, we take the range of $\sqrt s$ as 760\,MeV$\le\sqrt{s}\le$800\,MeV. From Eqs.~(27) and (29), we obtain $(m_\rho+m_K)^2<Q^2<m_\tau^2$. Hence we take the range of $\sqrt {Q^2}$ from $(m_\rho+m_K)=1270$\,MeV to $m_\tau=$1770\,MeV. The differential CP asymmetry $A_{CP}$ depending on $Q^2$ and $s$ is displayed in Fig.~\ref{dif}, where we take central values of the parameters involved in the calculation. We can see that $A_{CP}$ varies from around $10^{-12}$ to around $10^{-14}$. The maximum differential CP violating asymmetry can reach $-(5.6^{+2.9}_{-1.7})\times10^{-12}$, where the errors come from the uncertainties of the CKM matrix elements, the $\rho$-$\omega$ mixing parameters and the form factors in the leading order in $G_F$. As we expect, there is a peak for the CP violating parameter $A_{CP}$ when the invariant mass of the $\pi^+\pi^-$ pair is in the vicinity of the $\omega$ resonance for a certain $\sqrt{Q^2}$. $\rho$-$\omega$ mixing enlarges $A_{CP}$ by four orders of magnitude in some regions. Furthermore, we also find that the sign of $A_{CP}$ changes frequently in some regions of $Q^2$. This behaviour can be easily understood if one notes that the denominator of $A_1$ (and $B_1$), which is defined in Eq.~(19) [and (20)], changes its sign when $\sqrt{Q^2}$ crosses the pole. This will lead to cancellations when one performs the integration over $Q^2$ in some regions. These cancellations are found be be quite obvious around the peak when $\sqrt s=784\,\text{MeV}$. In the experimental study of three-body decays of the $B$ meson, one divides the Dalitz plot of candidates into bins with equal population using an adaptive binning algorithm and the CP violating parameter is calculated from the number of $B$ event candidates in each bin \cite{Aaij}. In the following, we will calculate the localized integrated CP violating parameter in the $\tau$ lepton decay, which may be measured in the future experiments. We will also compare CP asymmetries with and without $\rho$-$\omega$ mixing in the following.

Firstly, we preform the integration over $\sqrt {Q^2}$ while keeping $\sqrt s$ fixed. We divide the integration region into eight equal intervals: (1.30 GeV, 1.35 GeV), (1.35 GeV, 1.40 GeV), (1.40 GeV, 1.45 GeV), (1.45 GeV, 1.50 GeV), (1.50 GeV, 1.55 GeV), (1.55 GeV, 1.60 GeV), (1.60 GeV, 1.65 GeV) and (1.65 GeV, 1.70 GeV). In each interval, we integrate over $\sqrt{Q^2}$ and calculate the CP asymmetries with and without $\rho$-$\omega$ mixing. The results are denoted by $A^s_{CP}$ and shown in Fig.~5. Our numerical results show that the $\rho$-$\omega$ mixing mechanism enlarges the $A^s_{CP}$ by about 1-2 order when $\sqrt{s}$ is around $0.784\,\mathrm{MeV}$. All the three NP models in Ref.~\cite{Ken} predict that the $s$-depending differential CP asymmetries are to be order of $\mathcal{O} (10^{-3})$ when $\sqrt s$ varies in the region (0.760\,MeV, 0.800\,MeV). It can be seen from Fig.~5 that the maximum values of our results including $\rho$-$\omega$ mixing are about nine orders smaller than these predictions. Since the $\rho$-$\omega$ mixing mechanism provides an extra strong phase which enlarges the CP asymmetry, we suggest to measure the localized CP asymmetry in the region where the invariant mass of the $\pi^+\pi^-$ pair is around $780\,\text{MeV}$ in this decay process.

\begin{table}[htb]
\renewcommand{\captionlabeldelim}{.~}
\centering
\captionstyle{flushleft}
\setcaptionwidth{14cm}
\caption{The localized integrated asymmetries (in units of $10^{-12}$) with (without) $\rho$-$\omega$ mixing. The central values of the numerical results correspond to central values of the parameters involved in the calculation and the errors are from the uncertainties of the CKM matrix elements, the $\rho$-$\omega$ mixing parameters, and the form factors in the leading order in $G_F$.}
\renewcommand{\arraystretch}{1.5}
\begin{tabular*}{14cm}{@{\extracolsep{\fill}}cccc}
\hline
\hline
$\sqrt{Q^2}$\,(GeV)     &$A^{\Omega}_{CP}$  &   $\sqrt{Q^2}$\,(GeV)     &$A^{\Omega}_{CP}$ \\
\hline
(1.30, 1.35)   & $3.4^{+1.3}_{-2.6}$\ ($-0.30^{+0.07}_{-0.19}$)  &  (1.50, 1.55)   &  $-6.6^{+3.5}_{-4.1}$\ ($0.43^{+0.27}_{-0.30}$)     \\
\hline
(1.35, 1.40)   &$9.6^{+3.3}_{-4.9}$($0.093^{+0.053}_{-0.041}$)  &  (1.55, 1.60)   &  $-2.2^{+1.8}_{-0.9}$\ $(0.12^{+0.05}_{-0.06})$        \\
\hline
(1.40, 1.45)   &$63^{+24}_{-33}$\ ($0.013^{+0.008}_{-0.004}$)    &  (1.60, 1.65)   &  $-3.8^{+1.8}_{-2.2}$\ $(-82^{+47}_{-60})$     \\
\hline
(1.45, 1.50)   &$51^{+42}_{-16}$\ $(-0.20^{+0.03}_{-0.09})$     &  (1.65, 1.70)   &  $-3.4^{+2.2}_{-1.5}$\ $(-0.14^{+0.05}_{-0.01})$     \\
\hline
\hline
\end{tabular*}
\end{table}

Finally, we integrate $A_{CP}$ over both $\sqrt {Q^2}$ and $\sqrt{s}$ and obtain the localized integrated asymmetries $A^{\Omega}_{CP}$. Considering the significant region of $\rho$-$\omega$ mixing shown in Fig.~\ref{s}, we choose the integration interval of $\sqrt s$ to be from 0.775\,MeV to 0.795\,MeV. The numerical results of the localized integrated asymmetries with (without) $\rho$-$\omega$ mixing are shown in Table.~\ref{tab2}, where the central values of the numerical results correspond to central values of the parameters involved in the calculation and the errors are again from the uncertainties of the CKM matrix elements, the $\rho$-$\omega$ mixing parameters, and the form factors in the leading order in $G_F$. We can see in most of the intervals $\rho$-$\omega$ mixing enlarges the localized integrated asymmetries. The maximum increase is three orders of magnitude. These predictions lead to a new upper limit of the CP asymmetries based on the SM in this decay channel.

We have calculated the differential and integrated CP asymmetries within the SM in the $\tau^\pm\rightarrow K^\pm\pi^+\pi^-\nu_\tau$ decay taking $\rho$-$\omega$ mixing into account. It is worth noting that even the maximum localized integrated CP asymmetry, $6.3^{+2.4}_{-3.3}\times 10^{-11}$, is at least six orders smaller than that predicted based on NP in Ref.~\cite{Ken}. If any CP violation bigger than our predicted values is observed in experiments, one may consider the possibility of NP.

\section{Conclusion}

In the framework of the SM, CP violation in the $\tau$ lepton decay process arises from a nontrivial phase in the CKM matrix and is predicted to be zero in the leading order in $G_F$. However, Delepine pointed out that the CP-odd phase can arise from the second order weak process in the $\tau^\pm\rightarrow K^\pm\pi^0 \nu_\tau$ decay mode \cite{Delepine1}. Since $\rho$-$\omega$ mixing can provide very large CP asymmetries in some decay channels of heavy hadrons, we have tried to enlarge the CP asymmetry in the $\tau^-\rightarrow K^- \rho^0 (\omega)\nu_{\tau}\rightarrow K^- \pi^+\pi^-\nu_{\tau}$ decay via this mechanism.

We have first studied the differential CP asymmetry depending on $\sqrt{Q^2}$ and $\sqrt{s}$. The numerical results show that it varies from around $10^{-12}$ to around $10^{-14}$ and the maximum CP violating asymmetry can reach $-(5.6^{+2.9}_{-1.7})\times10^{-12}$. We have found that there is a peak for the CP violating parameter $A_{CP}$ when the invariant mass of the $\pi^+\pi^-$ pair is in the vicinity of the $\omega$ resonance. The advantage of $\rho$-$\omega$ mixing is that it makes the strong phase difference between the hadronic matrix elements of the leading order and the second order in $G_F$ larger at the $\omega$ resonance. Consequently, the CP violating asymmetry reaches the maximum value when the invariant mass of the $\pi^+\pi^-$ pair in the decay product is in the vicinity of the $\omega$ resonance. We have also found that $A_{CP}$ changes its sign when $\sqrt{Q^2}$ varies. Then, we have calculated the localized integrated CP violating parameter in the $\tau$ lepton decay, which may be measured in the future experiments.

After integrating over $\sqrt{Q^2}$ in serval intervals, we have shown that the $\rho$-$\omega$ mixing mechanism enlarges the $\sqrt s$ dependent CP asymmetry by about 1-2 order when $\sqrt{s}$ is around $0.784\,\mathrm{MeV}$. The maximum value of these results including $\rho$-$\omega$ mixing is about nine orders smaller than that predicted by NP model \cite{Ken}. Since the $\rho$-$\omega$ mixing mechanism provides an extra strong phase which enlarges the CP asymmetry, we suggest to measure the localized CP asymmetry in the region where the invariant mass of the $\pi^+\pi^-$ pair is around $780\,\text{MeV}$ in this decay process. Furthermore, the mechanism in present paper can also be considered in the $(3\pi)^\pm\nu_\tau$ and $(4\pi)^\pm\nu_\tau$ final states in the $\tau$ lepton decay.

At last, we also have preformed integration over both $\sqrt{Q^2}$ and $\sqrt{s}$ to obtain $A^{\Omega}_{CP}$. The maximum value of $A^{\Omega}_{CP}$ turns out to be $6.3^{+2.4}_{-3.3}\times 10^{-11}$. This value is the largest CP asymmetry in this decay channel within the SM predicted at present, which is still at least six orders smaller than that predicted based on NP in Ref.~\cite{Ken}, leaving plenty room for CP-violating NP to be discovered in the $\tau$ decay. If any differential CP asymmetry bigger than $10^{-12}$ or localized integrated CP violation bigger than $10^{-11}$ is observed in the future, it may be the signal of NP.

\end{spacing}
\end{document}